\documentclass[twocolumn,showpacs,preprintnumbers,superscriptaddress]{revtex4-1}
\usepackage{float}
\usepackage{amsmath}
\usepackage{amsfonts}
\usepackage{dcolumn}
\usepackage{bm}
\usepackage{epsfig}
\usepackage{epstopdf}
\usepackage{graphicx}
\usepackage{amsfonts}
\usepackage{amssymb}
\usepackage{mathrsfs}
\usepackage{color}
\usepackage{appendix}
\usepackage{multirow}
\usepackage{threeparttable}
\usepackage{verbatim}
\usepackage{graphicx,amsmath}
\usepackage{color,xcolor}
\usepackage[bookmarks=false]{hyperref}
\hypersetup{colorlinks=true,citecolor=blue,linkcolor=blue,urlcolor=blue,pdfstartview=FitH,bookmarksopen=true}

\begin{document}
	\title{Enhancement of Quantum Sensing in a Cavity Optomechanical System around Quantum Critical Point}
	\author{Shao-Bo Tang}
	\affiliation{School of Physics, Zhengzhou University, Zhengzhou 450001, China}
	\author{Hao Qin}
	\affiliation{School of Physics, Zhengzhou University, Zhengzhou 450001, China}
	\author{D.-Y. Wang}\email[]{dywang@zzu.edu.cn}
	\affiliation{School of Physics, Zhengzhou University, Zhengzhou 450001, China}
	\author{Kaifeng Cui}
	\affiliation{School of Physics, Zhengzhou University, Zhengzhou 450001, China}
	\author{S.-L. Su}\email[]{slsu@zzu.edu.cn}
	\affiliation{School of Physics, Zhengzhou University, Zhengzhou 450001, China}	
	\author{L.-L. Yan}\email[]{llyan@zzu.edu.cn}
	\affiliation{School of Physics, Zhengzhou University, Zhengzhou 450001, China}
	\author{Gang Chen}
	\affiliation{School of Physics, Zhengzhou University, Zhengzhou 450001, China}	
	\date{\today}
	
\begin{abstract}
The precision of quantum sensing could be improved by exploiting quantum phase transitions, where the physical quantity tends to diverge when the system approaches the quantum critical point. This critical enhancement phenomenon has been applied to the quantum Rabi model in a dynamic framework, showing a promising sensing enhancement without the need for complex initial state preparation. In this work, we present a quantum phase transition in the coupling cavity-mechanical oscillator system when the coupling strength crosses a critical point, determined by the effective detuning of cavity and frequency of mechanical mode. By utilizing this critical phenomenon, we obtain a prominent enhancement of quantum sensing, such as the position and momentum of the mechanical oscillator. This result provides an alternative method to enhance the quantum sensing of some physical quantities, such as mass, charge, and weak force, in a large mass system.
\end{abstract}
	
\maketitle
	
\section{INTRODUCTION}\label{sec1}
	
In the pursuit of checking new theories, a consistent imperative resides in the enhancement of measurement precision with the development of technologies~\cite{giovannettiQuantumEnhancedMeasurementsBeating2004a,giovannettiAdvancesQuantumMetrology2011a,degenQuantumSensing2017a,giovannettiQuantumMetrology2006b}. Quantum sensing, as one advancing technology to 		carry out the high-precision measurement, can significantly improve measurement precision by exploiting quantum phenomena such as quantum superposition, entanglement, and criticality \cite{sachdevQuantumPhaseTransitions2013a,ramsLimitsCriticalityBasedQuantum2018a,frerotQuantumCriticalMetrology2018a}. At the core of measurement precision lies the relevant parameters' susceptibility. By utilizing the high sensibility characteristic to minute changes in physical parameters around the quantum critical point, the measurement precision can be greatly enhanced\cite{liForceDependentQuantum2021,macieszczakDynamicalPhaseTransitions2016a,gietkaUnderstandingImprovingCritical2022,dicandiaCriticalParametricQuantum2023}. A study shows that the precision of coupling strength at the critical point exhibits a significant enhancement with respect to the non-critical ones, which establish a framework for quantifying quantum precision limits \cite{zanardiQuantumCriticalityResource2008a}.

Quantum sensing refers to the estimation of parameter values through an appropriate estimator following the measurement of a parameterized quantum state signal~\cite{mitchellSuperresolvingPhaseMeasurements2004,waltherBroglieWavelengthNonlocal2004,luisEquivalenceMacroscopicQuantum2001,jooQuantumMetrologyEntangled2011,vandamOptimalQuantumCircuits2007,macconeRobustStrategiesLossy2009,yurkeSUSUInterferometers1986,kimInfluenceDecorrelationHeisenberglimited1998}.The preparation of the initial state assumes paramount importance in this process of physics, because the inherent quality directly determines the precision of parameter estimation. Actually, one effective but challenging method to enhance the quantum sensing is to prepare a ground state close to the critical points. Recently, some works demonstrate the potential  for achieving high-precision measurement with a relaxed initial state preparation by utilizing quantum criticality \cite{chuDynamicFrameworkCriticalityEnhanced2021a,luCriticalQuantumSensing2022}. Quantum phase transition (QPT) manifests a significant divergence phenomenon, exemplified by the Dicke model \cite{dickeCoherenceSpontaneousRadiation1954a}, transitions from the normal phase to the superradiance phase in the thermodynamic limit. Further researches show that the QPT can also occur in a finite component system \cite{bakemeierQuantumPhaseTransition2012,garbeSuperradiantPhaseTransition2017,liuUniversalScalingCritical2017,chenFinitesizeScalingAnalysis2018}, such as the Rabi model \cite{rabiSpaceQuantizationGyrating1937}, where the QPT appears when the atomic transition frequency is much greater than optical field frequency \cite{ashhabSuperradianceTransitionSystem2013,hwangQuantumPhaseTransition2015,caiObservationQuantumPhase2021}. Subsequently, the QPT in the Rabi model has been proposed to enhance the quantum sensing for the precision measurement with relaxed initial state preparation \cite{chuDynamicFrameworkCriticalityEnhanced2021a,garbeCriticalQuantumMetrology2020,iliasCriticalityEnhancedQuantumSensing2022}. 

A typical cavity optomechanical system (COMS), consisting of a mechanical oscillator, optical cavity, and driving light~\cite{marquardtOptomechanics2009,verhagenQuantumcoherentCouplingMechanical2012,bowenQuantumOptomechanics2015,zhangPrecisionMeasurementElectrical2012,xiongAsymmetricOpticalTransmission2015}, is an emerging powerful platform to explore the nonlinear radiation-pressure coupling between mechanical mode and optical mode with important roles in precision measurement \cite{marquardtDynamicalMultistabilityInduced2006}. In past years, plenty of significant theoretical and experimental studies have been devoted to understanding the nonlinear dynamics in COMS. These studies have covered various aspects such as steady state dynamics with weak coupling, bistable state dynamics with strong coupling and some unstable  dynamics, including limit cycle, period  bifurcations and chaos\cite{zhuCavityOptomechanicalChaos2023a,ludwigOptomechanicalInstabilityQuantum2008a,zhuSinglephotontriggeredQuantumChaos2019,lorchLaserTheoryOptomechanics2014,bakemeierRouteChaosOptomechanics2015,schulzOptomechanicalMultistabilityQuantum2016,djorweFrequencyLockingControllable2018}. Recently, some important works have studied quantum metrology at the mesoscopic scale by nonlinear mechanical devices \cite{parisQuantifyingNonlinearityQuantum2014,asjadJointQuantumEstimation2023,tekluNonlinearityNonclassicalityNanomechanical2015,tekluNonlinearityNonclassicalityNanomechanical2015}. Yet, despite significant progress has been made in this field, the metrological analysis of quantum criticality in COMS still lacks substantial investigations. COMS exhibits a transition from stable state to unstable state  \cite{chenStrongSinglephotonOptomechanical2021,xiongOptomechanicalinterfaceinducedStrongSpinmagnon2023,liuDynamicalPhaseTransition2021,luSinglePhotonTriggeredQuantumPhase2018,jagerDynamicalPhaseTransitions2019,wangCollapseSuperradiantPhase2016}, which is an excellent platform to achieve enhancement of quantum sensing. 


In this work, we investigate the enhancement of quantum sensing by exploiting the QPT in a typical COMS. A divergent feature of the quantum Fisher information (QFI) is found near the quantum critical point, and it is subsequently elucidated that the proximity of the measurement precision achieved through quantum phase transitions (QPT) to the quantum Cram\'e-Rao bound is observed. \cite{fisherTheoryStatisticalEstimation1925,braunsteinStatisticalDistanceGeometry1994a,liuQuantumFisherInformation2014}. Moreover, the QFI does not require specific initial state preparation, thus simplifying the experimental requirements. Additionally, the adjustable drive laser in the COMS also provides a more convenient method to satisfy the frequency relationship than the standard Rabi model. Furthermore, the parameter of QPT is proportional to the optomechanical coupling strength, implying that the phase transition could be obtained by modulating the drive laser power. Since mechanical oscillators can be regarded as a bridge coupled to various physical systems and are suitable for detecting quantities such as mass, charge, and weak force \cite{barzanjehOptomechanicsQuantumTechnologies2022,sansaOptomechanicalMassSpectrometry2020,xiongHighlySensitiveOptical2017}, our work paves the way for critical quantum sensing by utilizing a mechanical oscillator around the quantum critical point.

The rest of this paper is organized as follows: In Sec.~\ref{sec2}, we provide the Hamiltonian of COMS in the rotating frame and then linearize it to discuss the dynamic phase transition around the quantum critical point. In Sec.~\ref{sec3}, we demonstrate the calculation process of the QFI for the parameter-dependent Hamiltonian. Sec.~\ref{sec4} shows the results of quadrature measurements by homodyne detection, and analyzes the error introduced by the finite frequency ratio with different initial states. Finally, we briefly discuss the experimental feasibility in Sec.~\ref{sec6} and give a conclusion in Sec.~\ref{sec7}.

\section{Quantum phase transition in COMS}\label{sec2}

\begin{figure}[htbh]
	\centering
	\includegraphics[width=1\linewidth]{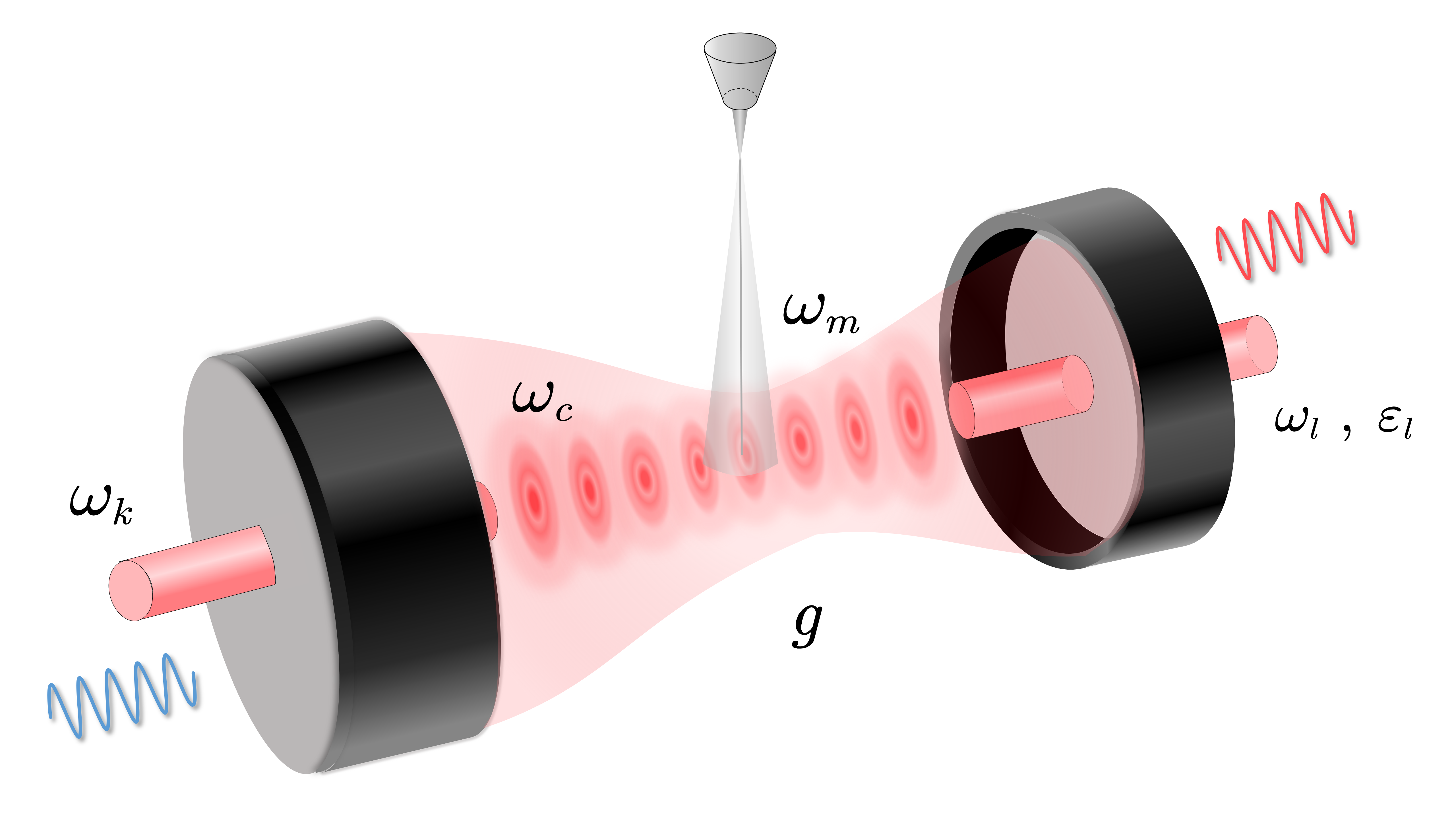}
	\caption{Cavity optomechanical system consists of an optical cavity with frequency $\omega_c$ (expressed by operators $a$ and $a^{\dagger}$), a mechanical oscillator with frequency $\omega_m$ (expressed by operators $b$ and $b^{\dagger}$) and $g$ is the optomechanical coupling strength. $\omega_k$ is the frequency of cooling laser and $\omega_l$ is the frequency of drive laser ($\varepsilon_{l}$ is amplitude).}
	\label{Fig1}
\end{figure}

A typical COMS is shown in Fig.~\ref{Fig1}, where the system consists of a single-mode field and a nanomechanical oscillator \cite{chanLaserCoolingNanomechanical2011a,foglianoMappingCavityOptomechanical2021}, and its Hamiltonian is ($\hbar=1$)
\begin{equation}
	\begin{aligned}
		H= & \ \omega_{c} a^{\dagger} a+ \omega_{m} b^{\dagger} b-g a^{\dagger} a\left(b^{\dagger}+b\right) \\ 
		&+i \varepsilon_{l}\left(e^{-i \omega_{l} t} a^{\dagger}-e^{i \omega_{l} t} a\right).
	\end{aligned}
	\label{mEq1}
\end{equation}

The first and second terms represent the free energy of cavity field and nanowire oscillator, the third term is the optomechanical coupling term with the coupling strength $g$, and the last term is regarded as the driving term with the driving amplitude $\varepsilon_l$.  Performing the rotation transformation under $U=e^{-i\omega_l t a^{\dagger} a }$ leads to the simplification of the Hamiltonian as follows: 
\begin{equation}
	H^{\prime} = \delta a^{\dagger} a+ \omega_{m} b^{\dagger} b-g a^{\dagger} a\left(b^{\dagger}+b\right)+i \varepsilon_{l}\left(a^{\dagger}-a\right).
	\label{mEq2}
\end{equation}
where $\delta=\omega_c-\omega_l$ is the detuning between the cavity field frequency and the driving laser frequency.


Replace the operators with their average values plus fluctuations driven by strong laser, i.e., $a\to \langle a \rangle+\delta a$, $b\to \langle b \rangle+\delta b$. In the following, we choose an appropriate driving phase to make $\left (\langle a \rangle,\langle b \rangle \right)$ real, and replace the fluctuation $\left(\delta a , \delta b\right)$ by $\left(a,b\right)$ for convenience. By neglecting the nonlinear terms and retaining the linear interaction terms, the resulting effective Hamiltonian is given by (Appendix~\ref{AA}): 
\begin{align}
	H_{L}= \Delta a^{\dagger} a+ \omega_{m} b^{\dagger} b-G\left( a^{\dagger}+ a\right)\left(b^{\dagger}+b\right), \label{mEq7}
\end{align}
where the effective detuning between the cavity field frequency and the driving laser frequency is $\Delta=\delta-2g\langle b \rangle$, and the enhanced effective optomechanical coupling strength $G=g\langle a \rangle$ \cite{wangSteadystateMechanicalSqueezing2016,wangSteadystateMechanicalSqueezing2016a}.

By utilizing the Schriffer-Wolff (SW) transformation \cite{schriefferRelationAndersonKondo1966,bravyiSchriefferWolffTransformation2011} to eliminate the off-diagonal part in $H_L$, we obtain an approximated appropriate unitary transformation $e^S$. The generator $S=G(b^\dagger+b)(a^\dagger-a)/ \omega_m$ is anti-Hermitian and block-off-diagonal under the condition of $\eta=\Delta/\omega_m \gg 1$, which makes the transformed Hamiltonian $e^{-S} H_L e^S$ free of the coupling terms between cavity field and nanomechanical oscillator. Consequently, generating the cavity field fluctuation to $\langle a^{\dagger}a\rangle_{(t=0)}=0$, we obtain the effective Hamiltonian as
\begin{equation}
	\mathcal{H}_L= \omega_m b^\dagger b -\frac{\lambda^2\omega_m}{4}(b^\dagger+b)^2,\label{mEq8}
\end{equation}
\begin{figure}
	\centering
	\includegraphics[width=1\linewidth]{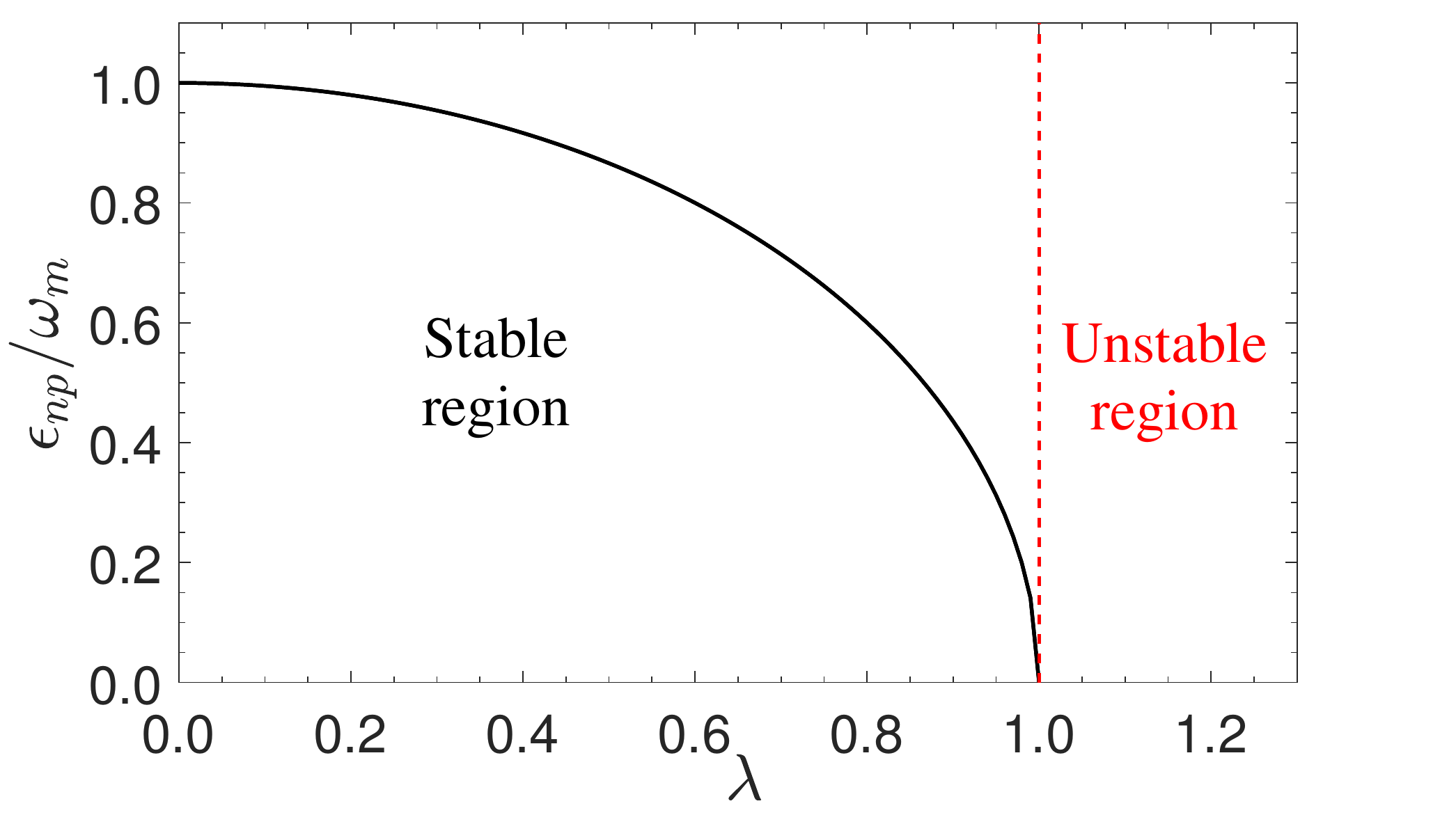}
	\caption{Variation  of  the  excitation frequency   with  respect  to  the  coupling  parameter. $\lambda < 1$ is the stable region we discussed, and $\lambda > 1$ enter the unstable region.}
	\label{Fig2}
\end{figure}
with $\lambda=2G/\sqrt{\Delta\omega_m}$. Eq.~(\ref{mEq8}) can be diagonalized by the squeezing operator $\mathcal{S}(r_{\rm np})=\exp[r_{np}(b^{\dagger 2}-b^2)/2]$ with squeezing amplitude $r_{\rm np}=\ln(1-\lambda^2)/4$, i.e., $\mathcal{H}^{\prime}_L=\mathcal{S}^{\dagger}(r_{\rm np})\mathcal{H}_L\mathcal{S}(r_{\rm np})=\epsilon_{\rm np}b^\dagger b+E_{\rm np}$ where $E_{\rm np}=(\epsilon_{np}-\omega_m)/2$ is the ground state energy and $\epsilon_{np}=\omega_m\sqrt{1-\lambda^{2} }$ is the excitation frequency \cite{hwangQuantumPhaseTransition2015}. Generally, the excitation frequency $\epsilon_{\rm np}$ is required to be real, which indicates $\lambda < 1$, i.e., the critical point is $\lambda_c =1$. If $\lambda < 1$, this system will be in stable state. While $\lambda > 1$, the excitation energy eigenvalues of the Hamiltonian turn into complex and the above processes is out of place, as shown in Fig.~\ref{Fig2}. Essentially, the dynamical evolution of COMS enters into unstable dynamics \cite{dasInstabilitiesUltrastrongCoupling2023,bakemeierRouteChaosOptomechanics2015,madiotBichromaticSynchronizedChaos2021}.

\section{QFI OF CRITICAL QUANTUM DYNAMICS}\label{sec3}

The performance of quantum sensing is determined by how sensitive two nearby states are, as indicated by QFI. Consider a parameter ($\xi$)-dependent Hamiltonian $H_\xi=H_0+\xi H_1$ and the system is initially prepared in $|\Psi_0\rangle$. After the dynamic evolution $U_\xi=\exp(-iH_\xi t)$, the state of the system becomes $\rho_{\xi}=U_{\xi}|\Psi_0\rangle\langle\Psi_0|U_{\xi}^\dagger$. The QFI, which characterizes the sensitivity between $\rho_\xi$ and $\rho_{\xi+\mathrm{d}\xi}$ with $\mathrm{d}\xi$ denoting the infinitesimal change of $\xi$, is $I_\xi=4\mathrm{\mathrm{Var}}[h_\xi]_{|\Psi_0\rangle}$, where $\mathrm{Var}[\dots]_{\left| \Psi_0 \right \rangle}$ represents the variance corresponding to the initial state, and $h_\xi=iU^\dagger_\xi\left(\partial_\xi U_\xi\right)$ is the generator of parameter translation with respect to $\xi$ \cite{pangOptimalAdaptiveControl2017,pangQuantumMetrologyGeneral2014a}. Therefore, the key mission to obtain QFI is calculating $h_\xi$. 

By the integral formula for the derivative of an operator exponential $\partial_x e^{-y H(x)}=-\int^{y}_{0}e^{-\left(y-s\right)H(x)}\partial_x H(x)e^{-s H(x)}\mathrm{d}s$ \cite{wilcoxExponentialOperatorsParameter1967}, we obtain the integral form of $h_\xi$ is
\begin{align}
	h_\xi &=\int^{t}_{0}e^{i H_\xi s} H_1 e^{-i H_\xi s}\mathrm{d}s= \int^{t}_{0}\sum_{n=0}^{\infty}\frac{(is)^n}{n!}\left[H_\xi^{(n)},H_1\right]\mathrm{d}s \notag \\
	&=-i\sum_{n=0}^{\infty}\frac{(it)^{n+1}}{(n+1)!}\left[H_\xi^{(n)},H_1\right]. \label{mEq9}
\end{align}
To further simplify the result of Eq.~(\ref{mEq9}) into two parts: linear term with time and oscillatory term with time, we first construct the Hamiltonian $H_\xi$ satisfying the following reciprocal relationship \cite{pangQuantumMetrologyGeneral2014a}
\begin{equation}
	\left [H_\xi,\Gamma \right]=\sqrt{\Lambda}\Gamma,\label{mEq10}
\end{equation}
where $\Lambda$ is dependent on the parameter $\lambda$, $H_\xi$ can be regarded as superoperator acting on $\Gamma$, and $\Gamma=i\sqrt{\Lambda}C-D$ with $C=-i\left[H_0,H_1\right]$ and $D=-\left[H_\xi,\left[H_0,H_1\right]\right]$. Then, making the system satisfy the following commutation rules 
\begin{equation}
	\begin{aligned}
		\left[H_\xi^{(2n+1)},H_1\right]&=i\Lambda^n C, \\
		\left[H_\xi^{(2n+2)},H_1\right]&=-\Lambda^n D, \quad \left(n \in \mathbb{N}\right),
	\end{aligned}
	\label{mEq11}
\end{equation}
Eq.~(\ref{mEq9}) can be further simplified to 
\begin{equation}
	h_\xi = H_1 t+\frac{\cos(\sqrt{\Lambda}t)-1}{\Lambda}C-\frac{\sin(\sqrt{\Lambda}t)-\sqrt{\Lambda}t}{\Lambda^{\frac{3}{2}}}D,  \label{mEq12}
\end{equation}
which reveals that $h_\xi$ tends to divergent as $\Lambda \rightarrow 0$ under the condition of $\sqrt{\Lambda}t\simeq \mathcal{O}(1)$, demonstrating a critical effect. Substituting Eq.~(\ref{mEq12}) into $I_\xi=4\mathrm{\mathrm{Var}}[h_\xi]_{\left| \Psi_0 \right \rangle}$ and take the limit as $\lambda\rightarrow 1$, we finally obtain the QFI as (see Appendix~\ref{A.B} for details) \cite{chuDynamicFrameworkCriticalityEnhanced2021a},
\begin{equation}
	I_\xi \simeq 4\frac{[\sin(\sqrt{\Lambda}t)-\sqrt{\Lambda}t]^2}{\Lambda^3} \mathrm{\mathrm{Var}}\left[D\right]_{|\Psi_0\rangle}.\label{mEq13}
\end{equation}
$I_\xi$ diverges as $\Lambda^{-3}$ under the conditions of $\sqrt{\Lambda}\simeq \mathcal{O}(1)$ and $\mathrm{Var}\left[D\right]_{\left|\Psi_0\right \rangle} \ne 0$ for the pure initial state $\left|\Psi_0\right \rangle$.

Assuming that Eq.~(\ref{mEq8}) satisfies the relation in Eq.~(\ref{mEq10}), we define the quadrature operators as $X=(b^\dagger+b)/\sqrt{2}$ and $P=i(b^\dagger-b)/\sqrt{2}$ before calculating the QFI of $\mathcal{H}_L$. Then, $\mathcal{H}_L$ can be rewritten as 
\begin{equation}
	\mathcal{H}_L^\prime=\frac{\omega_m}{2}\left[P^2+\left(1-\lambda^2\right)X^2\right]. \label{mEq14}
\end{equation}
Comparing the form of Eq.~(\ref{mEq14}) with $H_\xi=H_0+\xi H_1$, we choose $H_0=\omega_m P^2/2$ and $H_1=\omega_m X^2/2$. Utilizing the method of Eq.~(\ref{mEq10}), we get $\Lambda=4\omega_m^2\xi$ with the parameter $\xi=1-\lambda^2$ and
\begin{equation}
	\begin{aligned}
		C&=\frac{\omega_m^2(XP+PX)}{2}, \\ 
		D&=\omega_m^3\left[P^2-\left(1-\lambda^2\right)X^2\right].
	\end{aligned}
	\label{mEq15}
\end{equation}
At this time we can obtain the QFI of $\mathcal{H}_L$ for the measurement of the parameter $\lambda$ as
\begin{align}
	I_\lambda(t)&=(\partial_\lambda\xi)^2 I_\xi(t)  \label{mEq16} \\
	&\simeq 16 \lambda^2\frac{[\sin(\sqrt{ \Lambda }\omega_m t)-\sqrt{\Lambda}\omega_m t]^2}{\Lambda^3} \mathrm{\mathrm{Var}}\left[P^2\right]_{\left|\varphi\right \rangle}, \notag 
\end{align}
where the parameter $\Lambda=4(1-\lambda^2)\ll 1$ and $\mathrm{Var}\left[D\right]_{|\Psi_0\rangle} \simeq \omega_m^6\mathrm{\mathrm{Var}}\left[P^2\right]_{|\Psi_0\rangle}$ as $\xi\rightarrow 0$ for the initial state $\left|\Psi_0\right \rangle$. It also shows that $I_\lambda(t)$ is divergent as $\xi\rightarrow 0$ (i.e., $\Lambda\rightarrow 0$), thus, the measurement precision can be enhanced around the quantum critical point. Next, We will present a feasible high-precision measurement approach and draw a comparison with QFI.

\section{The measurement precision of system}\label{sec4}

\subsection{Superposition State as the Initial State} \label{sec4a}

\begin{figure}[htbp]
\centering
\includegraphics[width=1\linewidth]{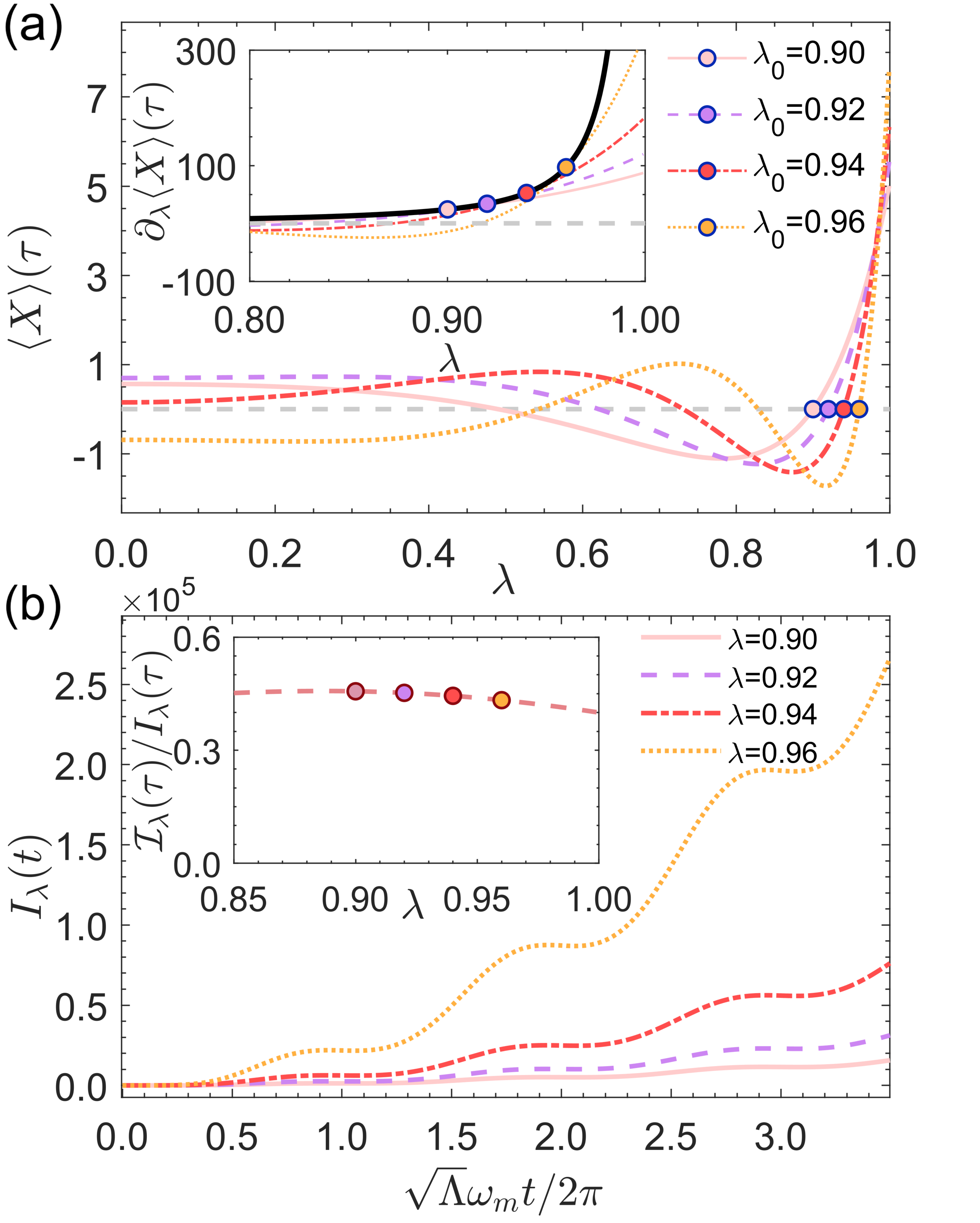}
\caption{Quantum sensing by homodyne detection of the mechanical field. (a) Quadrature $\langle X\rangle_{\tau} $ as a function of $\lambda$ after an evolution time $\tau=2 \pi/(\omega_m\sqrt{\Lambda_{\lambda_0}})$ with $\Lambda_{\lambda_{0}}=4(1-\lambda_0)^2$. The filled circles represent $\lambda=\lambda_0$. Inset: This figure shows the susceptibilty $\partial_{\lambda}\langle X\rangle_{\tau}$ as a function of $\lambda$ after the same evolution time $\tau$. The black solid line with filled circles illustrate the graphics of $\partial_{\lambda}\langle X\rangle_{\tau}$ at $\lambda_0=\lambda$. (b) The QFI $I_{\lambda}$ as a function of evolution time $t$. Inset: the ratio between $\mathcal{I}_{\lambda}(\tau)$ and $I_{\lambda}(\tau)$ after an evolution time $\tau=2 \pi/(\omega_m\sqrt{\Lambda})$.}\label{Fig3}
\end{figure}

Let us now assess the performances of standard homodyne detection with a product state $|\Psi_0\rangle=\left|0\right \rangle_{a}\otimes\left|\varphi\right\rangle_m$,  where the state of mechanical field mode $\left|\varphi\right\rangle_m=(\left|0\right\rangle+i\left|1\right\rangle)/\sqrt{2}$. There are two methods, namely classical fisher information and error propagation principle, to qualify the precision of quadrature measurement. In the main text, we employs the error propagation function to compare with QFI. After an evolution over a duration of time $t$, governed by the effective Hamiltonian of COMS in Eq.~(\ref{mEq14}), the motion equations of quadrature operators $X$ in the Heisenberg picture can be obtained as (see Appendix~\ref{A.C} for more details)
\begin{equation}
	\begin{aligned}
		&\langle X\rangle_{t}  =\sqrt{2} \Lambda^{-\frac{1}{2}} \sin \left(\sqrt{\Lambda} \omega_m t / 2\right),  \\
		&(\Delta X)^{2}  =1+\left( 2\lambda^2-1\right) \Lambda^{-1} \left[1-\cos \left(\sqrt{\Lambda} \omega_m t\right)\right]. 
	\end{aligned}
	\label{mEq17}
\end{equation}
As shown in Fig.~\ref{Fig3}(a), $\langle X\rangle_{t}$ as a function of $\lambda$ after an evolution time $\tau=\pi/\left[\omega_m\left(\sqrt{1-\lambda_0^2}\right)\right]$ is sensible around working point $\lambda_0$ and the derivative of $\langle X\rangle_{t}$ tends to diverge as $\lambda_0\rightarrow1$.

Defining a function $\mathcal{I}_\lambda(t)=(\partial_\lambda\langle X\rangle_{t})^2/(\Delta X)^2$\cite{giovannettiQuantumMetrology2006a,huelgaImprovementFrequencyStandards1997a}, the estimation error is then given by $\mathcal{I}^{-1}_\lambda(t)$. When $\mathcal{I}_\lambda(t)=I_\lambda(t)$, the estimation error reaches the quantum Cramér-Rao bound (i.e., reciprocal of QFI). As shown in Fig.~\ref{Fig3}(b), the periodic peaks of $\mathcal{I}_\lambda(t)$ are obtained at the evolution time $\tau_n=2n\pi/\sqrt{\Lambda}\omega_m$ with the peak value ${\mathcal{I} _{\lambda}}\left( \tau_n \right) =32\lambda ^2\pi ^2n^2{\Lambda}^{-3}$, which is the same order with the corresponding QFI ${I_{\lambda}}\left(\tau_n \right)\simeq64\lambda ^2\pi ^2n^2{\Lambda}^{-3} \mathrm{Var}\left[ P^2 \right] _{|\varphi \rangle _m}$ according to Eq.~(\ref{mEq16}), which can  been seen from the inset of Fig.~\ref{Fig3}(b). Note that the evolution time of $\mathcal{I}_\lambda(\tau_n)$ will increase in the vicinity of critical point because of $\tau_n \propto 1/\sqrt{\Lambda}$. However, it can be observed that both the values of $ I_{\lambda}\left(\tau_n \right) $ and ${\mathcal{I} _{\lambda}}\left( \tau_n \right)$ become larger as $\lambda\rightarrow 1$ and as increase of $\tau_n$. This indicates that the bound of precision is elevated (the larger $ I_{\lambda}\left(\tau_n \right) $), allowing for higher precision in measurements (the larger ${\mathcal{I} _{\lambda}}\left( \tau \right)$). Besides, our protocols do not require particular initial states of mechanical field mode. In other word, despite the presence of critical slowing down, the enhancement of quantum sensing can be realized without the need for a complex state preparation by encoding the physical parameter in COMS.

\subsection{The Influence of Finite Frequency Ratio} \label{sec4b}

The above discussion is based on the thermodynamic limit condition, i.e., $\eta=\Delta/\omega_m \rightarrow \infty$. This condition is idealized, used to neglects the higher order terms of the Hamiltonian of COMS after the SW transformation. however, $\eta$ can only attain finite values in the actual condition. Therefore, the purpose of this section is to explore the influence of higher-order terms in the SW transformation of the COMS Hamiltonian. After rectifying $S=G(b^\dagger+b)(a^\dagger-a)/ \omega_m$ into $\Tilde{S}$, we can derive the corrected Hamiltonian $\Tilde{H}_L$ (See Eq.~(\ref{ApdEq6}) and Eq.~(\ref{ApdEq7}) in Appendix~\ref{A.A}), which results in a correction on Eq.~(\ref{mEq17}), i.e., $ \langle \Tilde{X}\rangle_{t}= \langle X\rangle_{t}+ \langle X\rangle_{c}$ and $(\Delta \Tilde {X})^{2}=(\Delta X)^{2}+(\Delta X_{c})^{2}$. Hence, we can obtain the dynamics of the quadrature in the condition of finite frequency ratio as 
\begin{align}
	\langle \Tilde{X}\rangle_{t}&=\langle \Psi |e^{i H_L t} X e^{-i H_L t}|\Psi\rangle \notag \\
	&= \langle \Psi |e^{\Tilde{S}} e^{i \mathcal{H}_L t} e^{-\Tilde{S}} X e^{\Tilde{S}}e^{-i \mathcal{H}_L t}e^{-\Tilde{S}}|\Psi\rangle \notag \\
	&= \langle \Psi |\left[1+\mathcal{O}(\eta^{-\frac{1}{2}})\right]e^{i \mathcal{H}_L t} X \left[[1+\mathcal{O}(\eta^{-1})\right] \notag \\
	&\quad \quad e^{-i \mathcal{H}_L t}\left[[1+\mathcal{O}(\eta^{\frac{1}{2}})\right]|\Psi\rangle,\label{mEq18}
\end{align}
with  $|\Psi\rangle=|\Psi_0\rangle$. The leading term $\langle \Psi |e^{i \Tilde{H}_L t} X e^{-i \Tilde{H}_L t}|\Psi\rangle$ is equal to $\langle X\rangle_{t}$ and the dominant contribution to correction is on the order of $\eta^{-1/2}\langle X\rangle_{t} \sim \left(\eta \Lambda \right)^{-1/2}$.

\begin{figure}[htbp]
	\centering
	\includegraphics[width=1\linewidth]{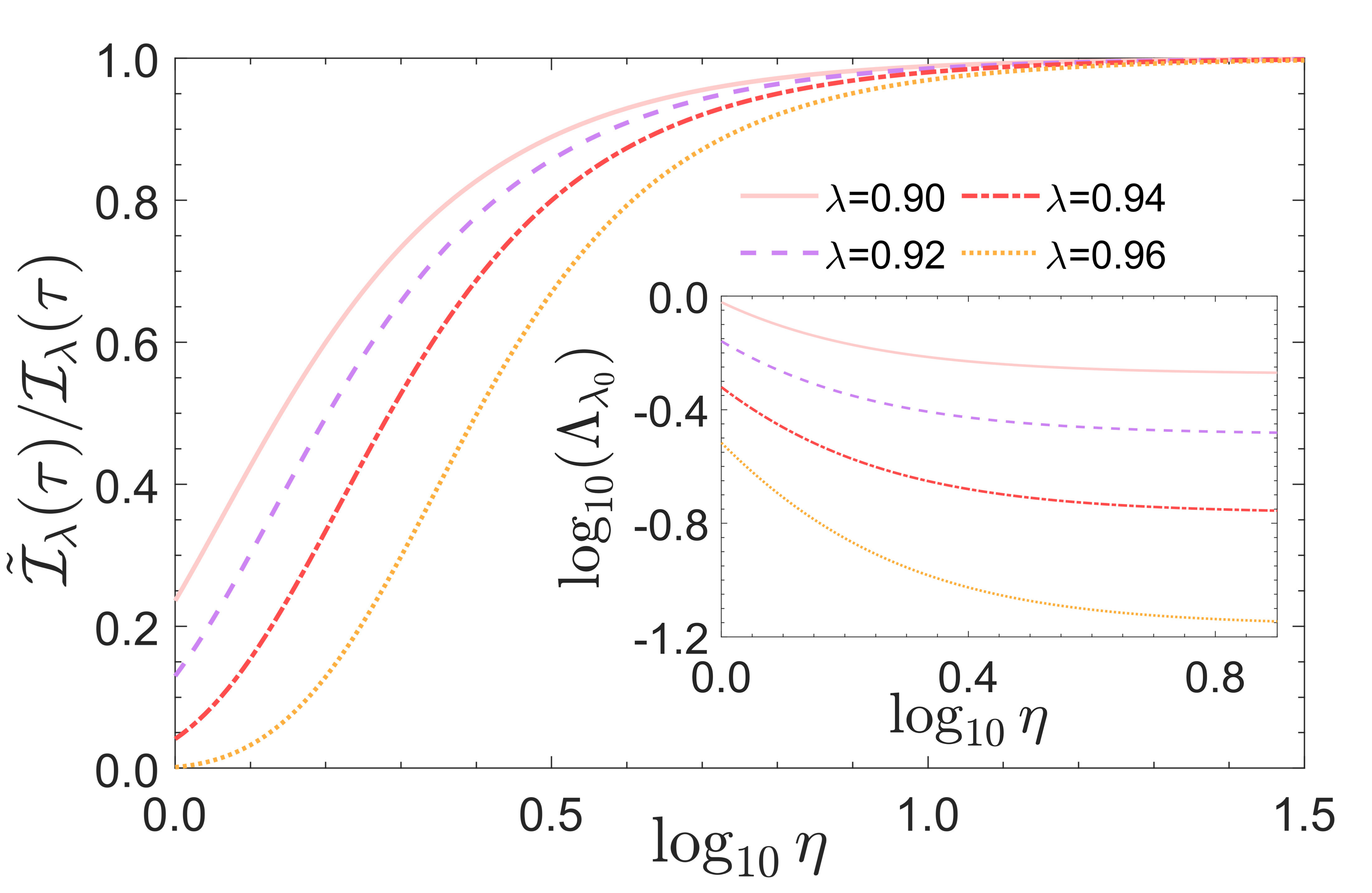}
	\caption{The ratio between $\mathcal{\Tilde{I}}_\lambda(\tau)$ and $\mathcal{I}_\lambda(\tau)$ for finite frequency ratio $\eta$ in the condition of $\eta\rightarrow\infty$ after an evolution time $\tau=2 \pi/(\omega_m\sqrt{\Lambda})$. Inset: the parameter $\Lambda_{\lambda_{0}}=4(1-\lambda_0)^2$ corresponds to the working point $\lambda_0$ at which $\mathcal{\Tilde{I}}_\lambda(\tau)$ achieves its local maximums. These solid curves satisfy the relation $\log_{10}[\Lambda_{\lambda_{0}}]=\log_{10}[4(1-\lambda)^2+\lambda^{4}\eta^{-2}]/3$.} \label{Fig4}
\end{figure}

In order to evaluate the influence of the transformed Hamiltonian $\Tilde{H}_L$, where the major influence is from $\lambda^2 \eta^{-1}$, we recalculate the Heisenberg equations of quadrature and obtain the correction $\langle X \rangle_c \sim \eta^{-2} \Lambda^{-3/2}$, which is more important than the order of $\left(\eta \Lambda \right)^{-1/2}$ derived from SW transformation. Similarly, the correction to the variance of the quadrature $(\Delta \Tilde{X})^2$ can be obtain as $(\Delta X_c)^2 \sim \eta^{-2} \Lambda^{-2}$. Thus, the analysis in the above section will remain valid in the condition of $\Lambda \gg \eta^{-1}$ and correction terms can be negligible compared with $\Delta X(\tau_n)=1$ when $t=\tau_n$. As shown in Fig.~\ref{Fig4}, the performance of our solutions can be sustained when the condition is satisfied.

\subsection{Measurement Precision of Coherent State}\label{sec5}

The above calculations are based on the mechanical superposition in Sec.~\ref{sec4}, a state which is not easy prepared in most mechanical systems. Here, we discuss the performance based on a coherence state around the critical point for more application. Firstly, we prepare a product state $\Psi_0=| 0 \rangle_\alpha \otimes |\alpha\rangle$, where $|\alpha\rangle$ denoting the coherence state of  mechanical oscillator. Then, the equations of quadrature operators $X$ can be obtained in the Heisenberg picture, with its mean value and variance given by
\begin{equation}
	\begin{aligned}
		&\langle X^{\left( \alpha \right)} \rangle \left( t \right) = \sqrt{2}\Lambda ^{-\frac{1}{2}}\mathrm{Im}\left( \alpha \right) \sin \left( \sqrt{\Lambda}\omega _mt/2 \right)  \\
		& \qquad \qquad \quad  +\sqrt{2}\mathrm{Re}\left( \alpha \right) \cos \left( \sqrt{\Lambda}\omega _mt/2 \right), \\
		&\left( \Delta X^{(\alpha)} \right) ^2\left( t \right) =\frac{1}{4}+\left( 1-\lambda^2 \right) \Lambda ^{-1}\cos(\sqrt{\Lambda}\omega t),
	\end{aligned}
	\label{mEq19}
\end{equation}
with Im($\alpha$) and Re($\alpha$) denoting the real part and imaginary part of $\alpha$, respectively. Thus, we have the function $\mathcal{I}^{(\alpha)}_{\lambda}(\tau_n)=(\partial_\lambda\langle X^{(\alpha)}\rangle_{\tau_n})^2/(\Delta X^{(\alpha)}_{\tau_n})^2=64\pi^{2}\lambda^2\Delta^{-3}\mathrm{Im}^2(\alpha)$ at evolution time $\tau_n=2n\pi/\sqrt{\Lambda}\omega_m$ with $\tau_n$ still denoting the positions of peak values of $\mathcal{I}^{(\alpha)}_{\lambda}(\tau_n)$. As expected, it keep the same order of ${I_{\lambda}^{(\alpha)}}\left(\tau_n \right)\simeq64\lambda ^2\pi ^2n^2{\Lambda}^{-3} \mathrm{Var}\left[ P^2 \right] _{|\varphi \rangle _m}$, which means that the coherence state can also promise the enhancement of quantum sensing. 

\begin{figure}[htbp]
	\centering
	\includegraphics[width=1\linewidth]{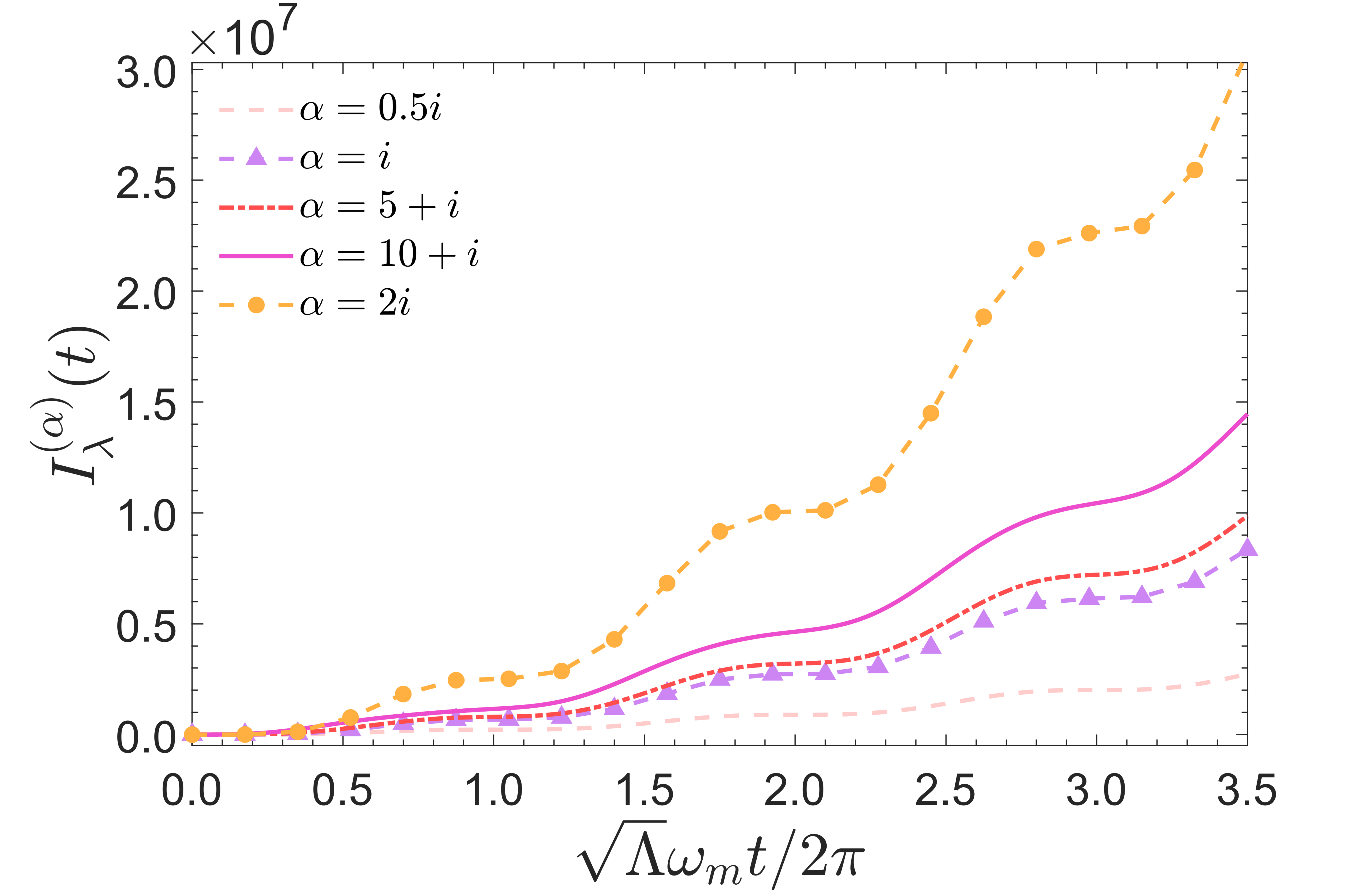}
	\caption{The QFI $I^{(\alpha)}_{\lambda}$ as a function of evolution time t with different $\alpha$, where we choose $\lambda=0.98$.} 
	\label{Fig5}
\end{figure}

\begin{figure}[htbp]
	\centering
	\includegraphics[width=1\linewidth]{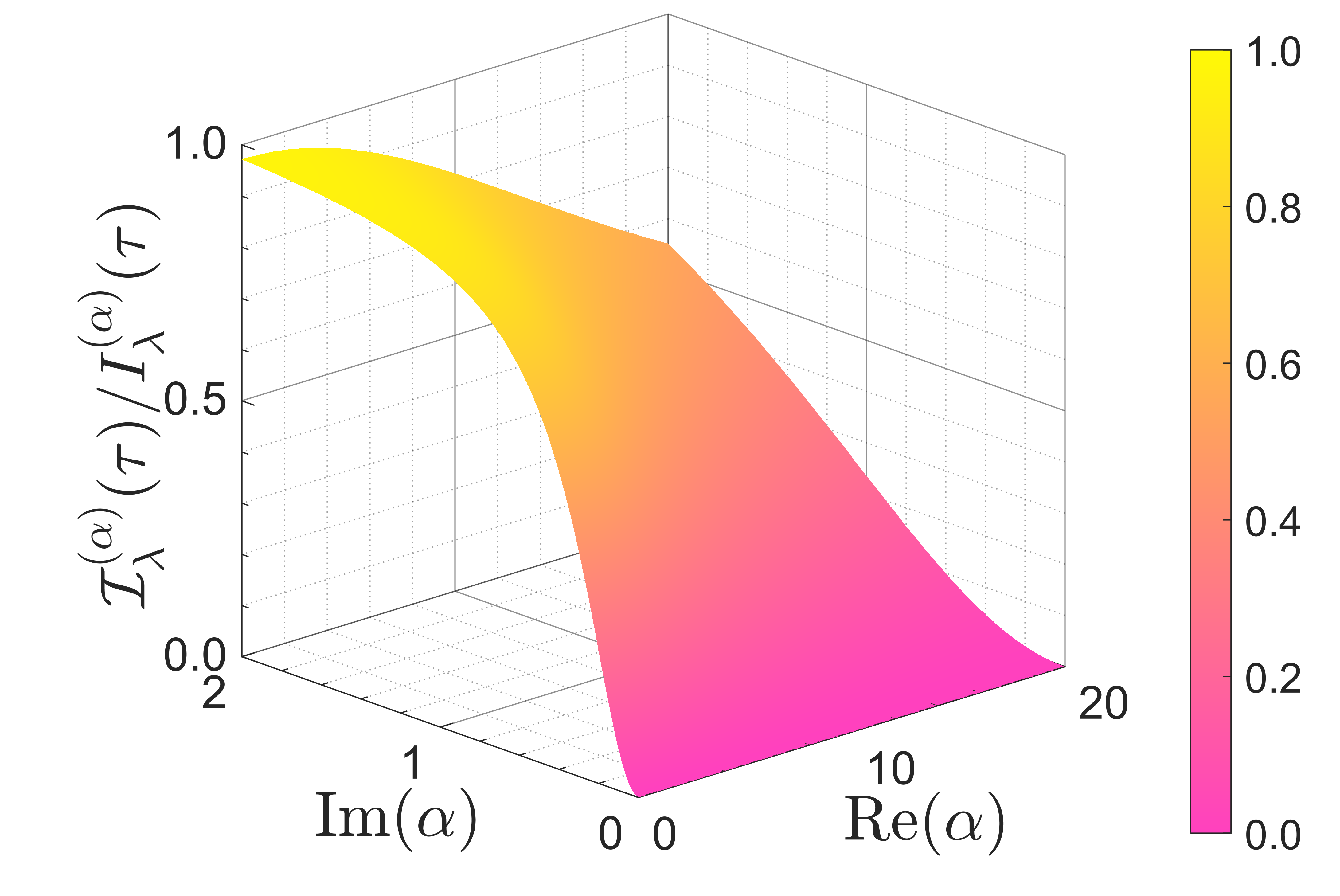}
	\caption{The ratio between $\mathcal{I}^{(\alpha)}_{\lambda}(\tau)$ and $I^{(\alpha)}{\lambda}(\tau)$ with $\lambda=0.98$ after an evolution time $\tau=\pi/(\omega_m\sqrt{1-\lambda^2})$. $\mathrm{Re}(\alpha)$ is negatively related to the ratio and $\mathrm{Im}(\alpha)$ is positively related to the ratio.}	\label{Fig6}
\end{figure}

Due to the similar critical behaviour as the superposition state, we assign $\lambda$ to $0.98$ to analyze the role of $\alpha$. As shown in Fig.~\ref{Fig5}, the QFI of the coherence state is affected by both $\mathrm{Re}(\alpha)$ and $\mathrm{Im}(\alpha)$. In addition, $\mathrm{Im}(\alpha)$ has a greater influence on $I_{\lambda}^{(\alpha)}$ and totally controls the value of $\mathcal{I}^{(\alpha)}_{\lambda}$ which can be reflected in Fig.~\ref{Fig6}. When $\mathrm{Im}(\alpha)$ takes a larger value, the estimation error is closer to Cram\'er-Rao bound. However, the ratio $\mathcal{I}^{(\alpha)}_{\lambda}/I_{\lambda}^{(\alpha)}$ diminishes as $\mathrm{Re}(\alpha)$ increases, which means that the real part of initial state can negatively impact the sensitivity of quantum sensing. The rationale behind this trend is that the function $\mathcal{I}^{(\alpha)}_{\lambda}$ remains constant while $I_{\lambda}^{(\alpha)}$ becomes larger, i.e., Cram\'er-Rao bound becomes lower. This critical behavior would own more superiority for the coherent state with a larger imaginary part. It is worth noting that the critical slowing down still exists, but more importantly, this example shows that different states (at least superposition state and coherent state) do not have a substantial affect on the ratio between function $\mathcal{I}_\lambda$ and QFI.  The enhancement of quantum sensing can be realized without harsh initial state preparation in COMS.

\begin{figure}[htbp]
	\centering
	\includegraphics[width=1\linewidth]{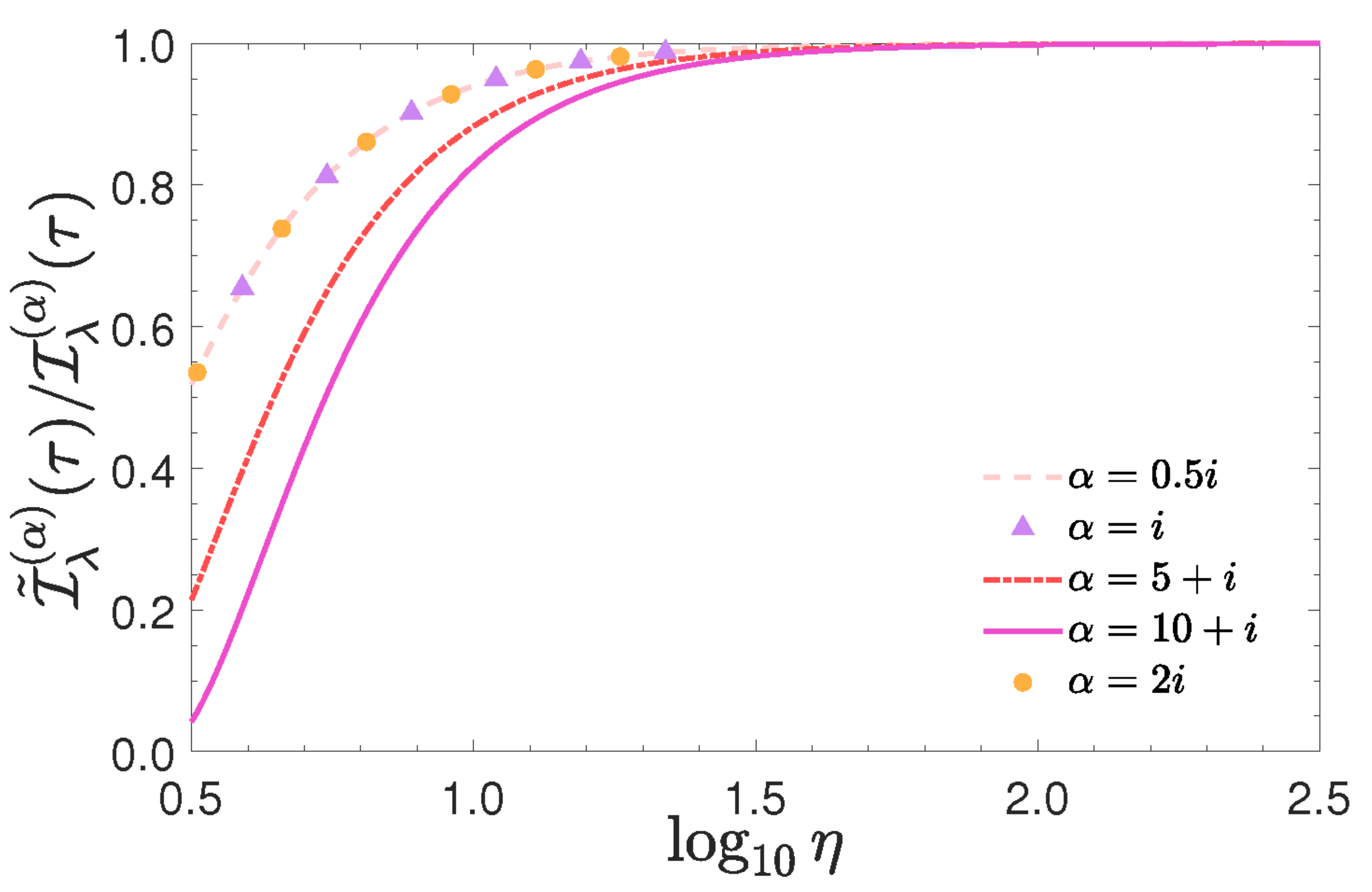}
	\caption{The ratio between $\mathcal{\Tilde{I}}^{(\alpha)}_\lambda(\tau)$ and $\mathcal{I}^{(\alpha)}_\lambda(\tau)$ for the finite frequency ratio $\eta$ in the condition $\eta\rightarrow\infty$ after an evolution time $\tau=\pi/(\omega_m\sqrt{1-\lambda^2})$ ($\lambda$=0.98). Dash line, triangle and circle are overlapped together when $\alpha$ takes a pure imaginary number.} 
	\label{Fig7}
\end{figure}

Next, we are curious about the performance of $\mathcal{I}^{(\alpha)}_{\lambda}$ with finite frequency. The major influence is still discussed by the motion equation of quadrature, $\langle X \rangle_c^{(\alpha)}\sim \eta^{-2} \Lambda^{-3/2}$ and $(\Delta X_c)^2 \sim \eta^{-2} \Lambda^{-2}$. When $\Lambda \gg \sqrt{2} \eta^{-1}$, the correction terms can be neglected. After calculation, we find that the ratio $\tilde{\mathcal{I}}^{(\alpha)}_{\lambda}/\mathcal{I}^{(\alpha)}_{\lambda}$ will not be affected by the imaginary part $\mathrm{Im}(\alpha)$ of coherent state. However, as shown in Fig.~\ref{Fig7}, the system requires a larger frequency ratio to reduce the impact of correction terms when $\mathrm{Re}(\alpha)$ increases.

\section{Experimental feasibility} \label{sec6}

The dynamic framework in this article requires less fluctuation of phonon number so that the QPT phenomenon of mechanical oscillator quantum fluctuation is evident. For this reason, we need to manipulate a strong laser to cool the nanomechanical oscillator into its ground state by some cooling methods, such as the sideband cooling \cite{teufelSidebandCoolingMicromechanical2011}, optical feedback cooling \cite{genesGroundstateCoolingMicromechanical2008}, and radiation pressure cooling \cite{arcizetRadiationpressureCoolingOptomechanical2006}. We insert a silicon carbide nanowire into a high-finesse fiber microcavity in static, insulated, and cryogenic ($T=15\ \text{mK}$) vacuum, where its quality factor is about $10^6$ and cavity length is  $200\ \mu\text{m}$ \cite{foglianoMappingCavityOptomechanical2021}. The employed nanowire with effective masses around $M=48\ \text{pg}$, diameter $d=130\ \text{nm}$, and length $L=10\ \mu\text{m}$ can vibrate with a frequency obout $\omega_m/2\pi = 10.56\ \text{MHz}$. To cool the nanomechanical oscillator, we can utilize an additional pump field with laser frequency $\omega_k$, satisfying the red-sideband resonance condition $\omega_c-\omega_k\sim\omega_m$. Moreover, the ground-state cooling of the nanomechanical oscillator has been realized experimentally in various systems \cite{petersonLaserCoolingMicromechanical2016,teufelOverwhelmingThermomechanicalMotion2016,chanLaserCoolingNanomechanical2011a, teufelSidebandCoolingMicromechanical2011}. The huge frequency ratio $(\omega_{c}-\omega_{l})/\omega_{m}$ needed in this proposal can be easily obtained by adjusting the frequency of driving laser $\omega_{l}$, where the frequency of driving laser and microcavity is in the optical range and much greater than the oscillator frequency. The optomechanical coupling strength is inherently tunable via the driving power, which further adjusts the parameter $\lambda$ near the quantum critical point. In addition, it is essential to prepare the nanomechanical oscillator into a superposition state or a coherence state~\cite{liaoGenerationMacroscopicSchrodingercat2016,hoffMeasurementInducedMacroscopicSuperposition2016,jahneCavityassistedSqueezingMechanical2009}. The realization of these states has already been reported in the trapped-ion oscillator system \cite{alonsoGenerationLargeCoherent2016,mccormickQuantumenhancedSensingSingleion2019}. In macroscopic ensemble, the coherent state has been realized in Kerr parametric oscillator \cite{frattiniSqueezedKerrOscillator2022}, and the superposition state has been relatively well established in theory \cite{abdiDissipativeOptomechanicalPreparation2016,liaoMacroscopicQuantumSuperposition2016,xieMacroscopicSuperpositionStates2019}. As a result, this dynamic framework is expected to be experimentally feasible and applied to super-precision measurement.

\section{CONCLUSION} \label{sec7}

In conclusion, we have investigated the transition of COMS and explored the quantum sensing around the quantum critical point . We demonstrate the feasibility of enhanced measurement by utilizing the critical divergent feature of QFI without the specific initial state preparation. Compared with the standard Rabi model, the huge frequency ratio between the cavity and the mechanical oscillator is adjustable and easy to be implemented by only changing the frequency of the driving laser in our proposal. Due to the excellent scalability of mechanical oscillators to various physical systems, the scheme in our work could be applied to quantum sensing for measuring classical quantities and macroscopic quantum phenomena.

\begin{acknowledgments}
This work was supported by the National Key Research $\&$ Development Program of China under Grant Nos. 2022YFA1404500, 2021YFA1400900, by National Natural Science Foundation of China under Grant Nos. 12274376, 12074232, 12125406, 12322410, U21A20434, 12074346, 12147149, 12204424, by the major science and technology project of Henan Province under Grant No. 221100210400, by the China Postdoctoral Science Foundation under Grant No. 2022M722889, by Natural Science Foundation of Henan Province under Grant Nos. 232300421075, 212300410085.
\end{acknowledgments}

\appendix

\renewcommand{\appendixname}{Appendix}	
\section{Linearization of the Optomechanical System}\label{AA}
Considering the influence of environmental thermal noise on COMS, the quantum langevin equation of Hamiltonian Eq.~(\ref{mEq2})  are given by
\begin{equation}
\begin{aligned}
\dot{a}&=\left(-i \delta-\frac{\gamma_c}{2}\right) a+i g a\left(b^{\dagger}+b\right)+\varepsilon_{l}+\sqrt{\gamma_c} a_{i n},  \\
\dot{b}&=\left(-i \omega_{m}-\frac{\gamma_{m}}{2}\right) b+i g a^{\dagger} a+\sqrt{\gamma_{m}} b_{i n},
\end{aligned} 
 \label{ApaEq1}
 \end{equation}
where $a_{in}$ and $b_{in}$ are noise operators of optical mode and mechanical mode, respectively, and $\gamma_{c,m}$ are the dissipation of system corresponding to the operators of $a_{in}$ and $b_{in}$.
	
Substitute the operators with their average values augmented by fluctuations induced by a strong laser, i.e., $a\to \langle a \rangle+\delta a$, $b\to \langle b \rangle+\delta b$. In the subsequent analysis, we replace the fluctuation $\delta a$ and $\delta b$ by $a$ and $b$, which is consistent with the main text. Then, the evolution equations for the mean amplitudes can be obtained as
\begin{equation}
    \begin{aligned}
		\dot{\langle a \rangle}&=\left[i\left(2g \langle b \rangle-\delta\right)-\frac{\gamma_c}{2}\right] \langle a \rangle+\varepsilon_{l},  \\
		\dot{\langle b \rangle}&=\left(-i \omega_{m}-\frac{\gamma_{m}}{2}\right) \langle b \rangle+i g{|\langle a \rangle|}^{2},
    \end{aligned}
    \label{Apaeq2}
\end{equation}
and the evolution equations of fluctuation operators are given as 
\begin{equation}
    \begin{aligned}
		\dot{a}&=\left[i\left(2g \langle b \rangle-\delta\right)-\frac{\gamma_c}{2}\right] a+i g \langle a \rangle \left(b^{\dagger}+b\right)+\sqrt{\gamma_c} a_{i n}, \\
		\dot{b}&=\left(-i \omega_{m}-\frac{\gamma_{m}}{2}\right) b+i g \langle a \rangle \left(a^{\dagger}+ a\right)+\sqrt{\gamma_{m}} b_{i n}.
    \end{aligned}
 \label{Apaeq3}
\end{equation}
Under a strong laser driving, the coherent amplitudes reach their steady state with 
\begin{equation}
\langle a \rangle=\frac{\varepsilon_l}{\gamma_c/2-i\left(2g \langle b \rangle-\delta\right)}, \ \
\langle b \rangle=\frac{ig{|\langle a \rangle|}^2}{i\omega_m+\gamma_m/2}.
\label{Apaeq4}
\end{equation}
At this time,  these Langevin equations of fluctuation operators is corresponding to the Hamiltonian Eq.~(\ref{mEq7}).

\section{Derivation of the QFI around Critical Point}\label{A.B}
The form of Eq.~(\ref{mEq13}) can be not directly obtained, but some terms can be discarded under some conditions \cite{chuDynamicFrameworkCriticalityEnhanced2021a}. Since the formula of $h_\xi$ has already been known, the exact QFI can be calculated as
\begin{equation}
	\begin{aligned}
		I_\xi&=4 t^2\mathrm{\mathrm{Var}}[H_1]_{\left|\Psi_0\right \rangle} + h^2(t)\Lambda^{-2} \mathrm{\mathrm{Var}}[C]_{\left|\Psi_0\right \rangle} \\
		&\quad+j^2(t)\Lambda^{-3}\mathrm{\mathrm{Var}}[D]_{\left|\Psi_0\right \rangle}+\mathcal{C}(t),\label{ApbEq1}
	\end{aligned}
\end{equation}
where $ h(t)=2[\sin(\sqrt{\Lambda}t)-\sqrt{\Lambda}t]$, $j(t)=2[\cos(\sqrt{\Lambda}t)-1]$, and the covariance term 
\begin{align}
	\mathcal{C}(t) = &2 h(t) t \Lambda^{-1} \operatorname{Cov}\left[H_{1}, C\right]_{|\Psi_0\rangle}-2 j(t) t\Lambda^{-\frac{3}{2}} \\
	&\operatorname{Cov}\left[H_{1}, D\right]_{|\Psi_0\rangle}-h(t) j(t) \Lambda^{-\frac{5}{2}} \operatorname{Cov}[C, D]_{|\Psi_0\rangle}, \notag \label{ApbEq2}
\end{align}
with the definition $\operatorname{Cov}\left[\mathcal{C}_1,\mathcal{C}_2\right]_{|\Psi_0\rangle}=\left\langle\Psi_0\left|\mathcal{C}_1 \mathcal{C}_2+\mathcal{C}_2 \mathcal{C}_1\right|\Psi_0\right\rangle-2\left\langle\Psi_0\left|\mathcal{C}_1\right| \Psi\right\rangle\left\langle\Psi_0\left|\mathcal{C}_2\right| \Psi_0\right\rangle$. Obviously, the third term of Eq.~(\ref{ApbEq1}) makes the dominant contribution as $\lambda \rightarrow 1$, i.e. the parameter $\xi \rightarrow \xi_c$. Thus, we can neglect the other low-order terms under the conditions $\sqrt{\Lambda}t=\mathcal{O}(1)$ and $\mathrm{\mathrm{Var}}\left[D\right] \ne 0$ which keep divergent scaling of $I_\xi$. The final approximate form is obtained as 
\begin{equation}
	I_\xi \simeq 4\frac{[\sin(\sqrt{\Lambda}t)-\sqrt{\Lambda}t]^2}{\Lambda^3} \mathrm{\mathrm{Var}}\left[D\right]_{\left|\Psi_0\right \rangle}.
	\label{ApbEq3}
\end{equation}

\section{Calculations of Quadrature Dynamic Evolution of COMS} \label{A.C}

In the Heisenberg picture, the equations of motion for the quadrature operators $X$, governed by Eq.~(\ref{mEq8}), are given as 
\begin{equation}
	\begin{aligned}
		\langle X\rangle _t&=\sqrt{2}\Lambda^{-\frac{1}{2}}\sin \left( \sqrt{\Lambda} \omega_m t/2 \right),
	\end{aligned}
	\label{ApcEq1}
\end{equation}
from which the susceptibility with respect to $\lambda$ can be obtained as
\begin{equation}
	\begin{aligned}
		\partial_{\lambda}\langle X \rangle_{t}=& 4 \sqrt{2} \lambda \Lambda^{-3 / 2} \sin \left(\sqrt{\Lambda} \omega_m t / 2\right) \\
		& -2 \sqrt{2} \lambda \omega_m t \Lambda^{-1} \cos \left(\sqrt{\Lambda} \omega_m t / 2\right). \label{ApcEq2}
	\end{aligned}
\end{equation}
Next, we calculate the expectation of the square operator in the Heisenberg picture as 
\begin{equation}
	\begin{aligned}
		\left< X^2 \right> _t=1+2\lambda ^2\Lambda^{-1}\left[ 1-\cos \left( \sqrt{\Lambda}\omega_m t \right) \right].\label{ApcEq3}
	\end{aligned}
\end{equation}
Then, we can obtain the variance of the quadrature $X$ and $P$ with the formula $(\Delta O)^2= \left< O^2 \right>_t- \left< O \right>^2_t$ as ($O$ denotes an operator)
\begin{equation}
	\begin{aligned}
		(\Delta X)^2&=1+\Lambda^{-1}\left( 2\lambda ^2-1 \right) \left[ 1-\cos \left( \sqrt{\Lambda} \omega_m t \right) \right].
	\end{aligned}
	\label{ApcEq4}
\end{equation}
According to $(\delta \lambda)^2=\left(\Delta O\right)^2/(\partial_\lambda\left<O\right>)^2 \ge 1/I_\lambda$ in quantum parameter estimation theory, we define $\mathcal{I}_\lambda=(\partial_\lambda\langle X\rangle_{t})^2/(\Delta X)^2 $. The significance of this function is that a greater $\mathcal{I}_\lambda$ value corresponds to higher precision, and as $\mathcal{I}_\lambda$ approaches $I_\lambda$, the precision approaches the limit of position measurement.

\begin{figure}[htbp]
	\centering
	\includegraphics[width=1\linewidth]{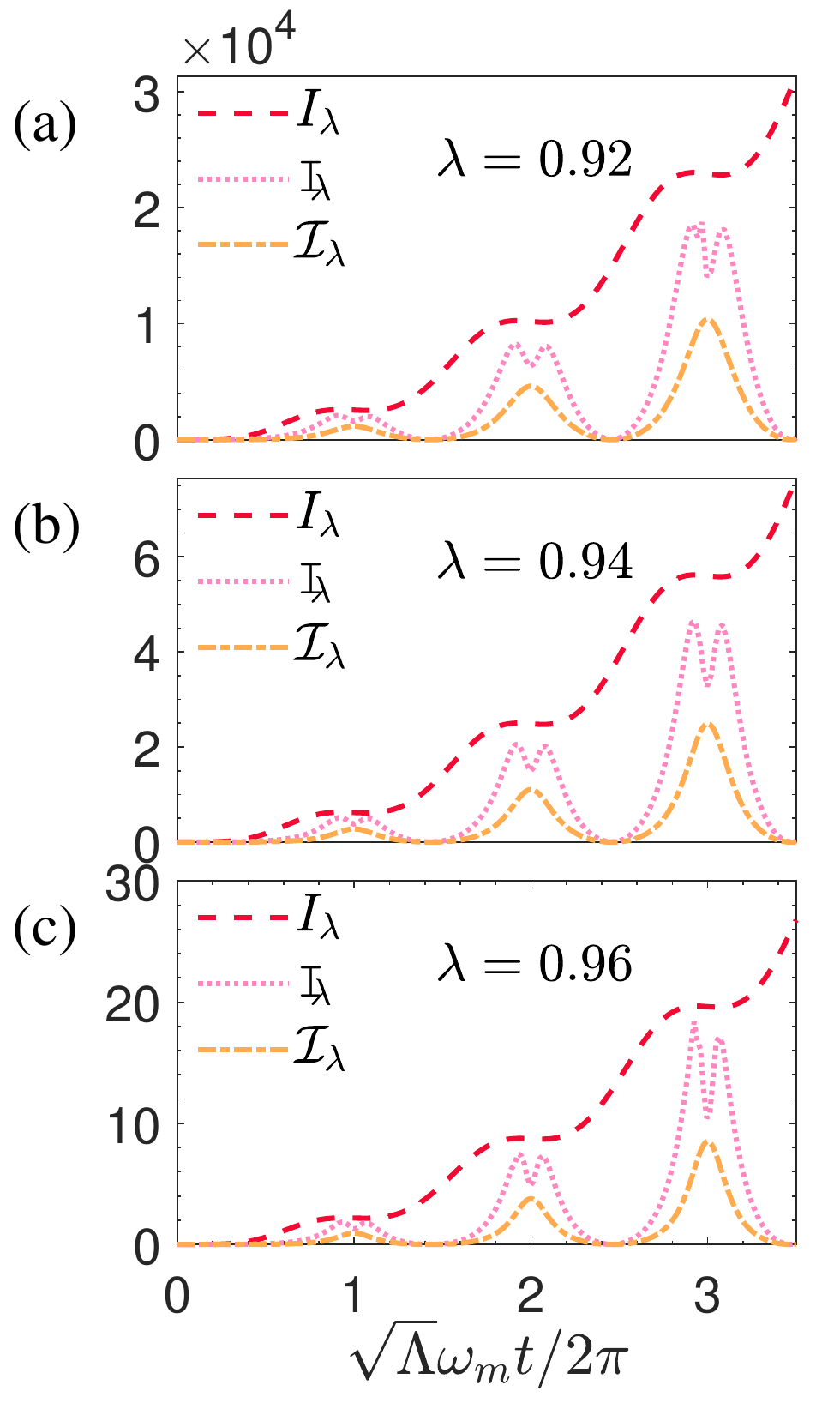}
	\caption{The $I_\lambda$ (Quantum fisher information), $\mathsf{I}_\lambda$ (Classical fisher information) and $\mathcal{I}_{\lambda}$ (Inverted parameter variance) for the quadrature operators $X$ as a function of $t$ with the periodic peaks occurring at even periods $\tau_n=2n\pi/(\sqrt{\Lambda}\omega_m)\left(n \in \mathbb{Z}^{+}\right)$. All of them increase with time $\tau_n$, which means a higher precision. The relative order remains $I_\lambda\geq\mathsf{I}_\lambda \geq \mathcal{I}_{\lambda} $, while $\lambda$ approaches the critical point.}
	\label{Fig8}
\end{figure}

Apart from inverted parameter variance $\mathsf{I}_\lambda$, Classical fisher information (CFI) can also quantifies the effectiveness of a certain observable.  Since $X$ has a continuous spectrum, the function of CFI is 
\begin{equation}
	\begin{aligned}
		\mathsf{I}_\lambda=\int dx\frac1{p(x|\lambda)}\biggl[\frac{\partial p(x|\lambda)}{\partial\lambda}\biggr]^2,
	\end{aligned}
	\label{ApcEq5}
\end{equation}
where $p(x|\lambda)=\mathrm{Tr}[\rho(\lambda)\Pi_{x}]$, the conditional probability distribution obtained from $X$ in state $\rho_\lambda(t)$. The POVM detector is obtained by $\Pi_{x}=\left|x\rangle \langle x\right|$. Besides, $|x\rangle$ requires transformation into the Fock representation by following formula:
\begin{equation}
	\begin{aligned}
	|x\rangle=\pi^{-1/4}\sum_n\frac{\exp(-x^2/2)}{(n!2^{n})^{1/2}}H_n(x)|n\rangle,
	\end{aligned}
	\label{ApcEq6}
\end{equation}
where $H_n(x)$ is the Hermite polynomial of order n. Additionally, the parameter error satisfies the inequality $(\delta \lambda^{(\text{CFI})})^2 \geq 1/\mathsf{I}_\lambda $ with observable operator $X$. Hence, one important relation can be given by
\begin{equation}
	\begin{aligned}
		I_\lambda\geq\mathsf{I}_\lambda \geq \mathcal{I}_{\lambda}.
	\end{aligned}
	\label{ApcEq7}
\end{equation}

The conditional probability distribution $p(x|\lambda)$ can be obtained through the numerical solution of $\rho_\lambda(t)$ at any time. Then, CFI $\mathsf{I}_\lambda$ can be computed using Eq.~(\ref{ApcEq5}), and a comparative analysis can be conducted with $I_\lambda$ and $\mathcal{I}$ illustrated in Fig.~\ref{Fig8}.

Firstly, The relationship of Eq.~(\ref{ApcEq7}) is verified from the graph. To ensure more precision assessment, we employ $\mathcal{I}_\lambda$ as the standard for qualify the precision of quadrature measurement in the main text. Additionally,  $\mathcal{I}(\tau_n)_\lambda=32\lambda ^2\pi ^2n^2{\Lambda}^{-3}$ has the same order with the QFI ${I_{\lambda}}\left(\tau_n \right)=64\lambda ^2\pi ^2n^2{\Lambda}^{-3} \mathrm{Var}\left[ P^2 \right] _{|\varphi \rangle _m}$ at local maximums point $t=\tau_n=2 n \pi/(\sqrt{\Lambda} \omega_{m})$. The detailed periodic change of $\mathcal{I}_\lambda$ can be witnessed from Fig.~\ref{Fig8}.

The figure of $\mathcal{I}_\lambda$ and $\mathsf{I}_\lambda$ can be obtained in the same way when the initial state is prepared in the coherence state. Here we will not repeat the calculation process. In brief, COMS model exhibits the same high precision at $\eta \rightarrow \infty$ compared with Rabi model. Additionally, this dynamic framework is a nice way to study mechanical cavity by bosonic field. 

\section{The SW Transformation of COMS with Higher-order Term}\label{A.A}

In the main text, the generator $S$ of SW transformation has already been approximated in the condition $\eta=\Delta/\omega_m \gg 1$. Here, we will reach a more precise generator $\Tilde{S}$ which holds for the finite frequency ratio \cite{hwangQuantumPhaseTransition2015,schriefferRelationAndersonKondo1966,bravyiSchriefferWolffTransformation2011}. The Hamiltonian $H_L$ satisfies the form of $H_s-G V$, with $H_s=\Delta a^{\dagger} a+\omega_m b^{\dagger} b$ and $V=(a^{\dagger}+a)(b^{\dagger}+b)$. Next, considering a unitary transformation $\Tilde{U}=e^{\Tilde{S}}$, the transformed Hamiltonian can be written as 
\begin{equation}
	\Tilde{H}_L=e^{-\Tilde{S}}H_L e^{\Tilde{S}}=\sum^\infty_{k=0} \frac{1}{k!}\left[H_L^{(k)},\Tilde{S}\right].\label{ApdEq1}
\end{equation}
Divide the transformed Hamiltonian into the diagonal part $\Tilde{H}_{\rm d}$ and off-diagonal part $\Tilde{H}_{\rm od}$ by defining $\Tilde{S}$ as block-off-diagonal and using the fact that $V$ is block-diagonal, which can be obtained as
\begin{equation}
	\begin{aligned}
		\Tilde{H}_d &= \sum^\infty_{k=0} \frac{\left[H_s^{(2k)},\Tilde{S}\right]}{2k!}-\sum^\infty_{k=0} \frac{\left[GV^{(2k+1)},\Tilde{S}\right]}{\left(2k+1\right)!},  \\
		\Tilde{H}_{od} &= \sum^\infty_{k=0} \frac{\left[H_s^{(2k+1)},\Tilde{S}\right]}{(2k+1)!}-\sum^\infty_{k=0} \frac{\left[GV^{(2k)},\Tilde{S}\right]}{\left(2k\right)!}. 
	\end{aligned}
	\label{ApdEq2}
\end{equation}
Now, we can give the generator $\Tilde{S}$ as 
\begin{align}
	\Tilde{S}=G\Tilde{S}_1+G^3\Tilde{S}_3 \label{ApdEq3}
\end{align}
by keeping to the third order in $G$.
$\Tilde{S}_1$ and $\Tilde{S}_3$ can be calculated by the formula 
\begin{equation}
	\begin{aligned}
		\left[H_s,\Tilde{S}_1\right]&=V, \\
		\left[H_s,\Tilde{S}_3\right]&=\frac{1}{3} \left[\left[V,\Tilde{S}_1\right],\Tilde{S}_1\right]. 
	\end{aligned}
	\label{ApdEq4}
\end{equation}
We find that the generator $\Tilde{S}$ satisfies the conditions in Eq.~(\ref{ApdEq4}) as
\begin{align}
	\Tilde{S}_1&=\frac{1}{\Delta}\left( b+b^{\dagger} \right) \left( a^{\dagger}-a \right) +\frac{\omega _m}{\Delta^2}\left( b-b^{\dagger} \right) \left( a+a^{\dagger} \right) \notag \\
	&\quad +\mathcal{O} \left( \frac{\omega _m}{{\Delta}^3} \right), \notag \\
	\Tilde{S}_3&=\frac{2}{3{\Delta}^3}\left[ \left( b+b^{\dagger} \right) ^3-\left( b-b^{\dagger} \right) ^2-\left( b+b^{\dagger} \right) \right] \left( a+a^{\dagger} \right) \notag \\ 
	&\quad +\frac{2\omega _m}{{3\Delta}^4}\left[ \left( a+a^{\dagger} \right) ^3-\left( a-a^{\dagger} \right) ^2-\left( a+a^{\dagger} \right) \right] \left( b+b^{\dagger} \right)\notag\\
	&\quad+\mathcal{O} \left( \frac{\omega _m}{{\Delta}^4} \right). \label{ApdEq5}
\end{align}
Next, we can use Eq.~(\ref{ApdEq3}) to rearrange the generator $\Tilde{S}$ as 
\begin{align}
	\Tilde{S}&=\frac{\lambda}{2}\eta^{-\frac{1}{2}}\left(b+b^\dagger\right)\left( a^{\dagger}-a \right) +\frac{\lambda}{2}\eta^{-\frac{3}{2}}\left( b-b^{\dagger} \right) \left( a+a^{\dagger} \right)\notag \\
	&\quad+\frac{\lambda^3}{12}\eta^{-\frac{3}{2}}\left[ \left( b+b^{\dagger} \right) ^3-\left( b-b^{\dagger} \right) ^2-\left( b+b^{\dagger} \right) \right] \left( a+a^{\dagger}\right) \notag \\
	&\quad+\mathcal{O}\left(\lambda^3\eta^{-\frac{5}{2}}\right),  \label{ApdEq6}
\end{align}
where $\lambda=2G/ \sqrt{\Delta\omega_m}$ and $\eta=\Delta/\omega_m$. Inserting the generator $\Tilde{S}$ into the Hamiltonian, we can reach
\begin{equation}
	\begin{aligned}
		\Tilde{H}_L&=\Delta a^\dagger a+ \omega_m b^\dagger b-\frac{\lambda^2}{4} \omega_m\left(b^\dagger+b\right)^2 \\
		&-\frac{\lambda^2}{4}\eta^{-1}\omega_m\left(a^\dagger+a\right)^2-\frac{1}{24}\lambda^4\eta^{-2}\omega_m \\
		&\left(b^\dagger+b\right)^2+\mathcal{O}\left(\lambda^4\eta^{-3}\right). \label{ApdEq7}
	\end{aligned}
\end{equation}
In the main text, Eq.~(\ref{mEq8}) only takes the first three terms for $\eta \gg 1$. When we consider the effect of finite frequency ratio of COMS, we cannot be neglect higher-order terms.

\bibliography{reference_b}

\end{document}